\documentclass[preprint2]{emulateapj}

\usepackage{natbib}  
\usepackage{amssymb}
\usepackage{multirow}

\usepackage{graphicx}
\usepackage{enumerate}
\usepackage{amsmath}
\usepackage{enumitem}

\def\cm{\,{\rm cm}}

\def\ergscm2 {erg\,s$^{-1}$cm$^{-2}$}

\def\cm2 {cm$^{-2}$}

\def\aap {A\&A}
\def\apj {ApJ}

\shorttitle{Specific predictions for SN 2014J in the context of the Quark Nova Ia model}
\shortauthors{Ouyed et al.}

\begin{document}
 
\title{Specific predictions for SN 2014J in the context of the Quark Nova Ia model}

\author{Rachid Ouyed\thanks{Email:rouyed@ucalgary.ca}, Nico Koning, Denis Leahy}
\affil{Department of Physics and Astronomy, University of Calgary, 2500 University Drive NW, Calgary, Alberta, T2N 1N4 Canada}

\author{Jan E. Staff}
\affil{Department of Physics and Astronomy, Macquarie University NSW 2109, Australia}

\begin{abstract}

The exact mechanism behind Type Ia supernovae (SNe-Ia) and the nature of the progenitors is poorly understood, although several theories vie for supremacy.  Their secured importance to the study of cosmology necessitates a resolution to this problem. Observations of nearby SNe-Ia are therefore critical to determine which theory, if any, is the correct one.  SN 2014J discovered at a relatively close 3.5 Mpc in the galaxy M82 provides such an opportunity.  In this paper we give specific predictions for SN 2014J in the context of the Quark Nova Ia (QN-Ia) model.  Predictions include X-ray luminosities just prior and hundreds of days after the explosion, light curve ``glitches",  neutrino emission, heavy element creation, and  gravitational signatures.
\end{abstract}



\section{Introduction}

Type Ia Supernovae (SNe-Ia) are thought to be ``standardizable" candles capable of measuring cosmological distances.  Their utility in this sense has been used over the past two decades to first discover (Riess et al. 1998; Perlmutter et al. 1999) and then study the accelerated expansion of the Universe; the results from which have deeply shaken our perception of nature.  Despite their overwhelming acceptance as ``standardizable" candles, the mechanism and nature of the progenitors behind the SN-Ia explosion are poorly understood.  We understand from their spectra that SNe-Ia are explosions of carbon/oxygen white dwarfs (COWDs); how they actually explode is still an area of intense research.  

Two theories lead the fray, both requiring the accumulation of mass onto the white dwarf (WD) leading to a thermonuclear explosion: the single-degenerate (SD) channel is the accretion onto the WD from a non-degenerate companion (Whelan \& Iben 1973) and the double-degenerate (DD) channel provides the mass from the merger of two WDs (Iben \& Tutukov 1994; Webbink 1984).  In the past year a new channel for the explosion of a WD has been presented (Ouyed \& Staff 2013), involving the explosion of a companion neutron star (NS) as a Quark-Nova Ia (QN-Ia).  Interestingly SNe-Ia formed through this channel are not standardizable using usual methods (Ouyed et al. 2013a) leading to profound implications for cosmology if the majority are formed in this way.

Resolving the progenitor/mechanism debate is therefore of the utmost importance for SNe-Ia to become more accurate tools for measuring cosmological distances, or even for determining if they can be used for such purposes at all.  Each of the aforementioned theories has unique signatures, that if observed would provide strong evidence for their cause.  However, SNe-Ia are difficult to study because they are rare and hence usually discovered at large distances.  The detection of nearby SNe-Ia therefore offers a unique opportunity to study these explosions in exquisite detail and perhaps lead to a resolution of the progenitor/mechanism debate.

The closest modern-era SNe-Ia observed prior to 2014 were SN 1972E in NGC 5253 at $\sim$ 2.5-8 Mpc  (e.g., Phillips et al. 1992; Sandage \& Tammann 1975;  della Valle \& Melnick 1992; Branch et al. 1994; Sandage et al. 1994) and SN 1986G in NGC 5128 at $\sim$ 3-5 Mpc.  Unfortunately at the time of these SNe, we did not have the space telescope capabilities we have today.  SN 2014J (Goobar et al. 2014) was discovered on Jan 21 2014 by astronomer Stephen J. Fossey in M82 at $\sim$ 3.5 Mpc, making it one of the closest SNe-Ia observed in the modern era.  With many of the most sophisticated telescopes trained on this object, SN 2014J provides an unprecedented opportunity to shed light on the mechanism and nature of the progenitors behind SN-Ia explosions.

In the coming months and years we expect a deluge of papers on SN 2014J spanning the entire electromagnetic spectrum.  In this paper we present observable predictions for SN 2014J in the context of the QN-Ia model in an attempt to make sense of what may or may not be seen and to provide motivation for upcoming observing proposals.

\section{Quark-Nova Ia}

Ouyed \& Staff (2013) considered the scenario in which a NS and COWD form a tight binary system, where mass transfer from the WD to the NS would occur.  The accreted mass would drive the NS above the critical mass sustainable by neutron matter and it would undergo a Quark Nova (QN) explosion to a quark star\footnote{The compact remnant (the QS) is born as an aligned rotator (Ouyed et al. 2004; Ouyed et al. 2006).} (QS)  (Ouyed et al. 2002; Vogt et al. 2004; Niebergal et al. 2010; Ouyed et al. 2013b).   The QN ejects the outermost layers of the NS at relativistic speeds with a Lorentz factor $\Gamma_{\rm QN}\sim 10$.  On average $10^{-3}M_{\odot}$ of iron-rich and neutron-rich material is ejected during a QN (Ker\"anen et al. 2005) equalling about $\sim 10^{52}$ erg in kinetic energy. This ejecta hits the WD fractions of a second after the QN explosion, leading to the thermonuclear explosion of the WD; the QN-Ia.  The properties of the QN ejecta as it hits the WD have been presented in \S 2.3 in Ouyed \& Staff (2013).  This external triggering mechanism and the induced shock compression implies that even low-mass WDs (i.e. $<< 0.5M_{\odot}$) will explode in the QN-Ia model. 
  
A QN-Ia, in addition to the energy from the $^{56}$Ni decay, is also powered by spin-down energy of the newly born QS. This results in the QN-Ia obeying a Phillips-like (calibration) relation where the variation in luminosity is due to spin-down power (see \S 4 in Ouyed et al. 2013a). We also find the calibration relation to be redshift-dependent which means that SNe-Ia are not standard candles \footnote{If the majority of SNe-Ia are in fact QNe-Ia} (see Ouyed et al. 2013a) making their utility as distance indicators unreliable.

\section{Current QN-Ia signatures in SN 2014J}

If SN 2014J is a QN-Ia explosion, several unique signatures may have already been observed prior to and in the few weeks following its discovery.

\begin{enumerate}[leftmargin=*]

\item  The hyper-accretion rate onto the NS just prior to the QN explosion should generate temperatures high enough for strong neutrino emission.  A luminosity on the order $0.1M_\odot /{\rm week} \sim 10^{46}$-$10^{48}$ erg s$^{-1}$ in tens of MeV neutrinos would be expected. For an $E_{\nu}\sim 10^{47}$ erg s$^{-1}$ and $\sim 10$ MeV neutrinos this would correspond to a flux of $\sim 10\ \nu$s cm$^{-2}$ s$^{-1}$.
This is clearly much below IceCube sensitivity (Abassi et al. 2011) but worth mentioning here.

\item Just prior to the neutrino dominated hyper-accretion phase, we expect a brief accretion phase ($< 1$ day) set by the photon Eddington limit ($L_{\rm X}\sim 10^{38}$ erg s$^{-1}$).  In the case of SN 2014J this would
correspond to a flux of $\sim 7\times 10^{-14}$ erg cm$^{-2}$ s$^{-1}$ which is detectable by Chandra (http://cxc.harvard.edu/). 

\item The NS will spin-up to millisecond periods due to accretion from the WD.  If the viewing angle is fortuitous, and the surrounding electron density low enough, this will have been observed as a radio pulsar in the days prior to the QN-Ia explosion.

\item  {\it Gravitational wave} (GW) detectors should see signatures of two explosions, the QN explosion and the WD detonation (a fraction of a second apart). The QN GW signatures have been investigated in Staff et al.  (2012).

\end{enumerate}

\section{Future QN-Ia signatures in SN 2014J}

Many signatures of the QN-Ia are not evident until the explosion becomes transparent to radiation. The following are a list of unique signatures that might be observed in SN 2014J in the months and years to come.

\begin{enumerate}[leftmargin=*]

\item The light curve of SN 2014J is expected, in the QN-Ia model, to undergo three distinct phases (see Figure 2 in Ouyed et al. 2013a):  (i) Spin-down dominated which lasts tens of days;  (ii) ($^{56}$Ni + $^{56}$Co) decay dominated; (iii) A return to spin-down dominated emission starting a few hundred days after the explosion.  We therefore expect to see a ``glitch" in the light curve of SN 2014J a few hundred days after the explosion, assuming a fiducial spin period of $\sim$ 20 ms, as the main energy source changes from ($^{56}$Ni + $^{56}$Co) decay to spin-down.  If the QS later collapses into a black hole (BH), a second ``glitch" will be observed as the spin-down energy will suddenly be extinguished.\footnote{We should note that if the QS collapses into a BH during phase ii. the first ``glitch" will never happen}

\item At $\sim 1$ year after the explosion we estimate the spin-down luminosity to be $\sim 10^{43}$ erg s$^{-1}$ for a QS born with an initial period of $\sim 20$ ms and an initial magnetic field of $\sim 10^{15}$ G. Assuming  an X-ray efficiency of $\sim 1$\% this would correspond to a flux of $\sim 10^{-10}$ erg cm$^{-2}$ s$^{-1}$.
Thus the compact remnant in SN 2014J should appear as a bright X-ray source $\sim 1$ year after the explosion. The QS could also be a marginally detectable Fermi source (in the 10-100 GeV band; http://fermi.gsfc.nasa.gov ) if a $\gamma$-ray efficiency of $\sim 1$\% is assumed.

\item The QN explosion proper ejects a very dense, ultra-relativistic ejecta with mass $M_{\rm QN}\sim 10^{-3}M_{\odot}$ and Lorentz factor $\Gamma_{\rm QN}\sim10$. The portion of the QN ejecta that will impact the WD is $\sim (R_{\rm WD}^2/4 a^2)\times M_{\rm QN}\sim 10^{-4}M_{\odot}$ where $R_{\rm WD}$ is the WD radius and $a\sim 2R_{\rm WD}$ the binary separation when the WD starts to fill its Roche-Lobe. This means that at most $\sim 0.1 M_{\rm QN} \sim 10^{-4}M_{\odot}$ will collide with the WD while the rest of the QN ejecta expands freely.   Thus most of the QN ejecta with its $\sim 10^{52}$ erg of kinetic energy will expand freely outwards with unique implications if it couples to the surrounding environment. For example, the QN ejecta may carve out a superbubble that could reach out to a parsec in a few years assuming a typical number density of $\sim 1$ particle per cc in the surrounding environment prior to the explosion.  HST should be able to resolve such a superbubble.

\item The neutron- and  iron-rich QN ejecta was shown to be an ideal site for the nucleo-synthesis of heavy elements, in particular nuclei with atomic weight $A>130$  (Jaikumar et al. 2007).   Compared to the burnt WD material, these nuclear proxies should be distinguishable in the late spectrum of SN 2014J.  However we predict at most $\sim 10^{-5}M_{\odot}$ of $A>130$ radioactive material to be mixed with the burnt CO ejecta. This is far too small to be detectable by Fermi and Nustar at the distance of SN 2014J.

\item The QS is likely to be born as an aligned rotator (Ouyed et al. 2004; Ouyed et al. 2006) and as such no radio pulsar should be seen in future observations of SN 2014J.

\item We have argued in previous work that the QN compact remnant shows properties reminiscent of 
 Soft Gamma-Ray Repeaters (SGRs) and Anomalous X-ray Pulsars (AXPs)  (see
Ouyed et al. 2010 and references therein). We therefore expect magnetar-like behaviour  from the location of SN 2014J in the future.

\end{enumerate}

\section{Discussion \& Conclusion}

The proximity of SN 2014J offers us an unparalleled opportunity to study a SN-Ia  which may reveal clues as to the nature of the progenitors and the explosion mechanism.  The two leading explosion scenarios (SD and DD channels) have recently been joined by a new intriguing possibility; the QN-Ia.  

The relative unfamiliarity of the QN-Ia model makes it easy to dismiss.  However, the QN has been successfully applied to a plethora of other astronomical phenomena including SGRs and AXPs (e.g. Ouyed et al 2010), Gamma-ray bursts (e.g. Ouyed et al. 2011) and Superluminous-supernovae (e.g. Ouyed et al. 2012).  In fact the double-humped light curve observed in SN 2009ip and SN 2010mc (modelled and well fit by Ouyed et al. 2013c) was first predicted by the QN model in 2009 (Ouyed et al. 2009), four years prior to its discovery.  Successes put aside, the QN-Ia model does make bold claims that if true could once again alter our perception of nature.  It is therefore imperative that we either rule out this scenario, or  confirm it with observations from near-by SNe-Ia such as SN 2014J.

In this paper we have provided the observer with a list of QN-Ia signatures which, if observed in SN 2014J, would support the
existence of  QNe-Ia.  We do note, however, that many of the late-time predictions rely on the existence of a QS.  It is entirely possible that the QS collapses to a BH during the ($^{56}$Ni + $^{56}$Co) decay phase (see point 1 in \S 4) thereby providing an (albeit unintended) hedge.  If the QS-BH transition occurs at any other time, however, we should see this in the light curve of SN 2014J (e.g. if it occurs immediately, no spin-down energy will be deposited and the light curve will be purely due to ($^{56}$Ni + $^{56}$Co) decay).

\begin{acknowledgements}   

This research is supported by  operating grants from the National Science and Engineering Research Council of Canada (NSERC). N.K. would like to acknowledge support from the Killam Trusts.

\end{acknowledgements}


\end{document}